\documentclass[twocolumn,aps,showpacs,prb]{revtex4-1}

\usepackage{amsmath,amssymb,graphics,epsfig,epstopdf,color,multirow,array,verbatim,ulem,braket,tabularx}
\usepackage[colorlinks,linkcolor=blue,citecolor=blue,urlcolor=blue]{hyperref}
\usepackage{color}
\usepackage{graphicx}
\usepackage{epstopdf}
\usepackage{amsmath}
\usepackage{multirow}
\usepackage{ulem}

\begin{document}

\title{First-principles study of the superconductivity in LaO}

\author{Pei-Han Sun}
\author{Jian-Feng Zhang}
\author{Kai Liu}\email{kliu@ruc.edu.cn}
\author{Qiang Han}\email{hanqiang@ruc.edu.cn}
\author{Zhong-Yi Lu}\email{zlu@ruc.edu.cn}

\affiliation{Department of Physics and Beijing Key Laboratory of Opto-electronic Functional Materials $\&$ Micro-nano Devices, Renmin University of China, Beijing 100872, China}

\date{\today}

\begin{abstract}

A recent experiment reported the first rare-earth binary oxide superconductor LaO ($T_c $ $\sim$ 5 K) with a rock-salt structure [K. Kaminaga \textit{et al}., J. Am. Chem. Soc. {\bf140}, 6754 (2018)]. Correspondingly, the underlying superconducting mechanism in LaO needs theoretical elucidation. Based on first-principles calculations on the electronic structure, lattice dynamics, and electron-phonon coupling of LaO, we show that the superconducting pairing in LaO belongs to the conventional Bardeen-Cooper-Schrieffer (BCS) type. Remarkably, the electrons and phonons of the heavy La atoms, instead of those of the light O atoms, contribute most to the electron-phonon coupling. We further find that both the biaxial tensile strain and the pure electron doping can enhance the superconducting $T_c$ of LaO. With the synergistic effect of electron doping and tensile strain, the $T_c$ could be even higher, for example, 11.11 K at a doping of 0.2 electrons per formula unit and a tensile strain of $4\%$. Moreover, our calculations show that the superconductivity in LaO thin film remains down to the trilayer thickness with a $T_c$ of 1.4 K.

\end{abstract}

\pacs{}

\maketitle

\section{INTRODUCTION}

Searching for new superconductors and exploring the related superconducting mechanism have been a long-standing subject in condensed matter physics. Among various superconductors, the binary lanthanide compounds have been intensively investigated. For example, the lanthanum dicarbide LaC$_{2}$ was shown to become superconducting at 1.6 K \cite{tc1969}, and the recent experimental \cite{exp2014} and theoretical \cite{the2015} studies further confirmed its BCS character. Compared with LaC$_{2}$, the lanthanum sesquicarbide La$_{2}$C$_{3}$ has a superconducting $T_c$ of 11 K \cite{11tc1,11tc2,11tc3}, which can be further raised to 13 K by tuning the La/C ratio \cite{13tc1,13tc2,13tc3}. However, whether the pairing symmetry in La${_2}$C${_3}$ is of multiple gap \cite{multip1,multip2,multip3} or $s$-wave gap \cite{13tc3} needs further clarification \cite{review2017}. The lanthanum monopnictides LaSb and LaBi exhibit extremely large magnetoresistance (XMR) at ambient pressure \cite{XMR1,XMR2,XMR3,XMR4} and were reported to show superconducting behavior under pressure\cite{LaSbTc,LaBiTc}. Nevertheless, our first-principles calculations indicated that the electron-phonon coupling (EPC) mechanism cannot account for the observed superconductivity in compressed LaBi \cite{jfzhang2020}. More recently, the lanthanum superhydride LaH$_{10}$ under ultrahigh pressure demonstrates a very high $T_c$ of 250-260 K\cite{LaH10exp1,LaH10exp2}, which is consistent with previous predictions based on EPC calculations \cite{LaH10the1,LaH10the2}. The numerous lanthanide compounds as exemplified above have thus provided a material warehouse for discovering new superconductors.

In addition to the aforementioned binary lanthanide superconductors, many unconventional superconductors, such as cuprate and Fe-based superconductors, also contain lanthanide elementary units. For example, the cuprate La$_{2-x}$Ba$_{x}$CuO$_{4}$ \cite{LaBaCuO} is composed of CuO$_{2}$ planes intercalated with LaO planes \cite{CuO21,CuO22}, while the iron-based superconductor LaO$_{1-x}$F$_x$FeAs \cite{LaOFeAs} consists of FeAs and LaO layers \cite{FeAs1,FeAs2,FeAs3}. The LaO layers in these two compounds are recognized as charge reservoirs that can provide carriers to CuO$_{2}$ or FeAs planes \cite{CuO21,CuO22,FeAs2,FeAs3}. On the other hand, rare attention has been paid to the binary lanthanum compound LaO, probably due to its poor chemical stability in bulk form \cite{instablity}. With the advance in the state-of-the-art oxide thin-film epitaxy techniques, several rare-earth monoxides with rock-salt structure including LaO have been successfully synthesized in the past five years\cite{monoxide1,monoxide2,monoxide3,monoxide4,monoxide5}. Particularly, the LaO thin films with a thickness of approximately 20 nm deposited on different substrates show bulk superconductivity with a $T_c$ around 5 K \cite{monoxide4}. This was an experimental report on the superconductivity in the rare-earth binary oxide. By changing the amount of oxygen vacancies in LaO thin films, the electron carrier density and the superconducting $T_c$ can be effectively tuned; moreover, the superconducting $T_c$ of LaO thin films varies from 4.25 K to 5.24 K when grown on YAlO$_3$, LaAlO$_3$, LaSrAlO$_4$ substrates \cite{monoxide4}. However, the electronic structure and the origin of superconductivity in LaO wait for theoretical clarification.

To explore the superconducting mechanism in LaO, we carried out first-principles calculations on its electronic structure, lattice dynamics, and EPC strength. We find that LaO is a phonon-mediated superconductor. The calculated superconducting $T_c$ agrees quite well with the previously measured value. We then studied the effects of lattice strain, charge doping, and dimensionality on the superconductivity in LaO.

\section{COMPUTATIONAL DETAILS}

The electronic structure, phonon spectrum, and EPC strength of LaO were studied by using the density functional theory (DFT) \cite{dft1,dft2} and density functional perturbation theory (DFPT) \cite{dfpt1,dfpt2} calculations as implemented in the Quantum ESPRESSO (QE) package \cite{QE1}. The interactions between electrons and nuclei were described by the RRKJ-type ultrasoft pseudopotentials \cite{uspp} which were taken from the PSlibrary \cite{pslibrary1,pslibrary2}. The generalized gradient approximation (GGA) of Perdew-Burke-Ernzerhof (PBE) formula  \cite{PBE} was adopted for the exchange-correlation functional. The kinetic energy cutoff of the plane-wave basis was set to 80 Ry. A 12$\times$12$\times$12 {\bf k}-point grid was adopted for the Brillouin zone (BZ) sampling. The Gaussian smearing method with a width of 0.004 Ry was used for the Fermi surface broadening. Both lattice constants and internal atomic positions were optimized with the Broyden-Fletcher-Goldfarb-Shanno (BFGS) quasi-Newton algorithm \cite{BFGS} until the forces on all atoms were smaller than 0.0002 Ry/Bohr. In the calculations of the dynamical matrix and the EPC, the BZ was sampled with a 6$\times$6$\times$6 {\bf q}-point mesh and a dense 60$\times$60$\times$60 {\bf k}-point mesh, respectively. At certain conditions ($2\%$ strain and 0.2 e/cell doping), a denser 8$\times$8$\times$8 {\bf q}-point mesh was also used to examine whether or not the imaginary frequency exists near the BZ center. The ultrathin LaO films were studied by two-dimensional slabs with a 16-{\AA} vacuum layer to avoid the artificial interaction between periodic images. For the LaO slabs, all atomic positions and lattice constants were fully optimized. A 12$\times$12$\times$1 {\bf k}-point grid was used for the self-consistent calculation while 6$\times$6$\times$1 {\bf q}-point and 60$\times$60$\times$1 {\bf k}-point grids were adopted to investigate the phonon dispersion and the EPC.

Based on the EPC theory, the Eliashberg spectral function $\alpha^2F(\omega)$ is defined as \cite{Eliashberg}

\begin{equation}
\alpha^2F(\omega)=\frac{1}{2{\pi}N(\varepsilon_F)}\sum_{{\bf q}\nu}\delta(\omega-\omega_{{\bf q}\nu})\frac{\gamma_{{\bf q}\nu}}{\hbar\omega_{{\bf q}\nu}},
\end{equation}
where $N(\varepsilon_F)$ is the density of states (DOS) at Fermi level $\varepsilon_F$, $\omega_{{\bf q}\nu}$ is the frequency of the $\nu$-th phonon mode at wave vector {\bf q}, and $\gamma_{{\bf q}\nu}$ is the phonon linewidth \cite{Eliashberg},
\begin{equation}
\gamma_{{\bf q}\nu}=2\pi\omega_{{\bf q}\nu}\sum_{{\bf k}nn'}|g_{{\bf k+q}n',{\bf k}n}^{{\bf q}\nu}|^2\delta(\varepsilon_{{\bf k}n}-\varepsilon_F)\delta(\varepsilon_{{\bf k+q}n'}-\varepsilon_F),
\end{equation}
in which $g_{{\bf k+q}n',{\bf k}n}^{{\bf q}\nu}$ is the electron-phonon coupling matrix element. The total electron-phonon coupling constant $\lambda$ can be obtained by \cite{Eliashberg}
\begin{equation}
\lambda=\sum_{{\bf q}\nu}\lambda_{{\bf q}\nu}=2\int{\frac{\alpha^2F(\omega)}{\omega}d\omega}.
\end{equation}
Then the superconducting transition temperature $T_c$ can be determined by substituting the EPC constant $\lambda$ into the McMillan-Allen-Dynes formula \cite{McMillan1},
\begin{equation}
T_c=\frac{\omega_{log}}{1.2}\text{exp}[\frac{-1.04(1+\lambda)}{\lambda(1-0.62\mu^*)-\mu^*}],
\end{equation}
where $\mu^*$ is the effective screened Coulomb repulsion constant with the empirical values between 0.08 and 0.15\cite{miu1,miu2}. In our calculation, $\mu^*$ was set to 0.13, similar to a previous study on lanthanum monochalcogenides \cite{LaS}. $\omega_{log}$ is the logarithmic average of Eliashberg spectral function that is defined as
\begin{equation}
\omega_{log}=\text{exp}[\frac{2}{\lambda}\int{\frac{d\omega}{\omega}\alpha^2F(\omega){ln}(\omega)}].
\end{equation}

\begin{figure}[!t]
\includegraphics[angle=0,scale=0.43]{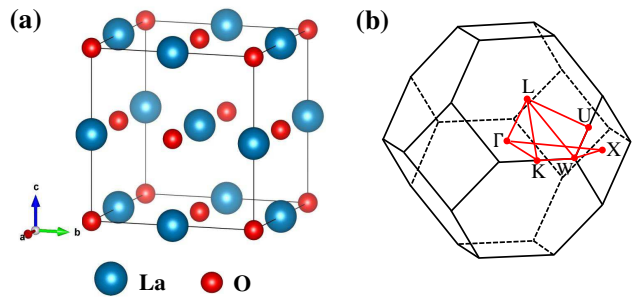}
\caption{(Color online) (a) Crystal structure of bulk LaO, where the blue and red balls represent La and O atoms, respectively. (b) Bulk Brillouin zone (BZ) for the primitive cell of LaO. The high-symmetry paths in BZ are indicated by red lines.}
\label{fig1}
\end{figure}

\section{Results and Analysis}

Figure \ref{fig1}(a) shows the bulk crystal structure of lanthanum monoxide LaO. Similar to other reported rare-earth monoxides \cite{monoxide1,monoxide2,monoxide3,monoxide5}, LaO adopts a rock-salt structure with space group Fm-3m, which is composed of two face-centered cubic lattices, respectively, of La and O. The calculated lattice constants of bulk LaO are $a=b=c=5.164$ {\AA}, in good accordance with the previous experimental values ($a=b=c=5.144$ {\AA})\cite{beforexp}. The bulk Brillouin zone (BZ) of the primitive cell of LaO along with the high-symmetry {\bf k} points is displayed in Fig. \ref{fig1}(b).

\begin{figure}[!t]
\includegraphics[angle=0,scale=0.53]{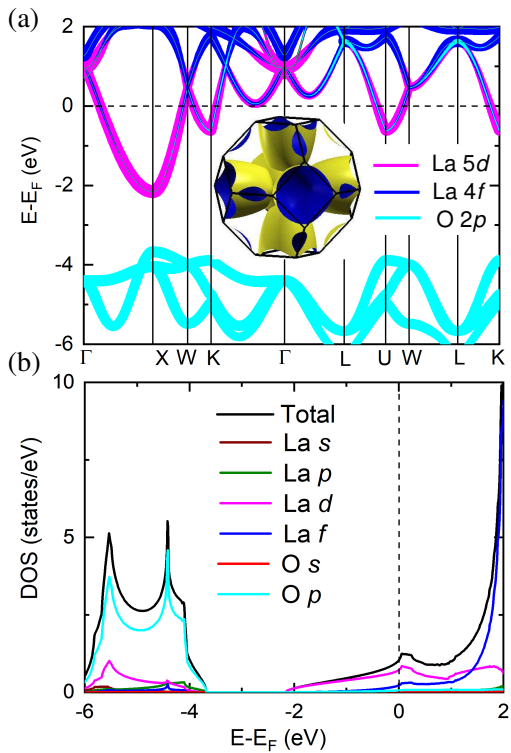}
\caption{(Color online) (a) Orbital-resolved electronic band structure of bulk LaO along the high-symmetry paths of BZ. The bands projected onto the La 5$d$, 4$f$, and O 2$p$ orbitals are displayed with pink, blue, and cyan lines, respectively. The widths of lines are proportional to the weights of the corresponding orbitals. Inset shows the Fermi surface of LaO. (b) Total and projected density of states of LaO. The Fermi level is set to zero.}
\label{fig2}
\end{figure}

Figure \ref{fig2} presents the orbital-resolved electronic band structure of bulk LaO along the high-symmetry paths in the BZ as well as the total and projected density of states (DOS). These results were calculated in the absence of spin-orbital coupling (SOC). There is a band passing through the Fermi level $E_F$ across the BZ [Fig. \ref{fig2}(a)], indicating the metallic behavior of LaO. Based on the orbital analysis, the 5$d$ orbitals of La play a dominant role around the Fermi energy level, followed by the 4$f$ orbitals, while the contribution of the 2$p$ orbitals of O is tiny. From the Fermi surface of LaO displayed in the inset of Fig. \ref{fig2}(a), we can see that there are large electron-type pockets around the X points and small electron-type pockets around the K and U points, respectively. The corresponding DOS shows a hump near the $E_F$ [Fig. \ref{fig2}(b)], which is helpful to the appearance of superconductivity. Moreover, the electronic states around the $E_F$ are mainly contributed by La 5$d$ orbitals, suggesting their major role playing in the superconductivity. This is consistent with the above analysis of orbital weights for the electronic bands. We also examined the electronic structure with the inclusion of SOC and found that both the energy dispersion and the total DOS near the Fermi level are almost unchanged. Hence, we will not consider the SOC effect in the following calculations.

\begin{figure}[!t]
\includegraphics[angle=0,scale=0.6]{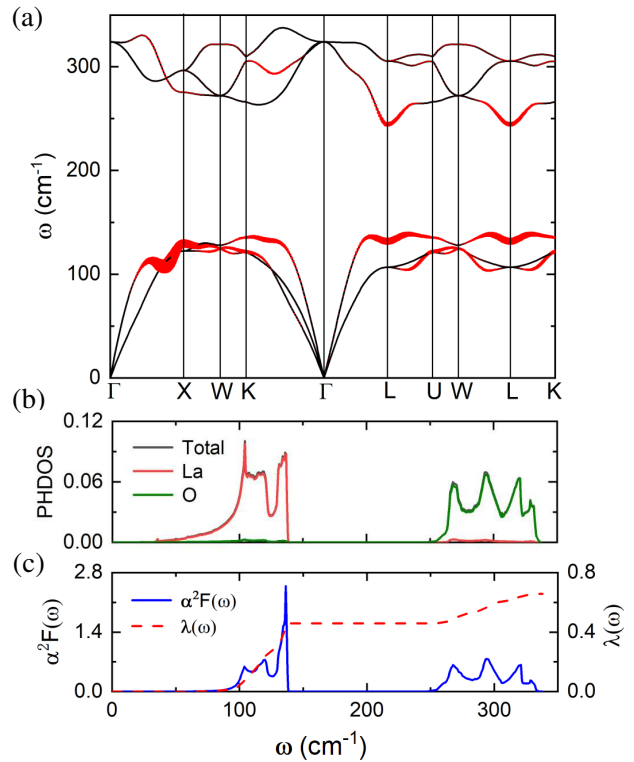}
\caption{(Color online) (a) Phonon spectrum of bulk LaO. The size of red dot represents the electron-phonon coupling (EPC) strength $\lambda_{{\bf q}\nu}$. (b) The total and projected phonon DOS. (c) Eliashberg spectral function $\alpha^2F(\omega)$ (blue line) and integrated EPC constant $\lambda(\omega)$ (red dashed line).}
\label{fig3}
\end{figure}

We next explore the phonon properties and the electron-phonon coupling of LaO. Figures \ref{fig3}(a) and \ref{fig3}(b) show the phonon dispersion with the EPC strength and the phonon density of states (PHDOS), respectively. There is no imaginary frequency in the phonon spectrum of Fig. \ref{fig3}(a), indicating the dynamical stability of bulk LaO. From the momentum- and mode-resolved EPC parameter $\lambda_{{\bf q}\nu}$ [as indicated by the size of red dot in Fig. \ref{fig3}(a)], we learn that the largest contribution to the EPC comes from the acoustic phonon branch in the frequency range of 100 to 140 cm$^{-1}$, which mainly originates from La vibrations [Fig. \ref{fig3}(b)] and contributes about $64.5\%$ to the total EPC constant [Fig. \ref{fig3}(c)]. In particular, the phonon modes with the frequencies $\sim$135 cm$^{-1}$ across the BZ almost all contribute significantly to the EPC [Fig. \ref{fig3}(a)], being consistent with the sharp peak in the Eliashberg spectral function $\alpha^2F(\omega)$ [Fig. \ref{fig3}(c)]. The calculated total EPC constant $\lambda$ of bulk LaO is 0.653. Based on the McMillan-Allen-Dynes formula [Eq. (4)], we obtained the superconducting $T_c$ of LaO as 4.84 K, which agrees well with the measured value of $\sim$5 K for LaO films grown on several substrates \cite{monoxide4}. This means that the superconductivity in rock-salt structure LaO can be explained in the framework of the BCS theory.

\begin{table}[!b]
\caption{Calculated lattice constants, electronic density of states, EPC parameters, and superconducting $T_c$s of bulk LaO under different biaxial strains $\epsilon$. $N(0)$ represents the electronic DOS at Fermi level. $\omega_{log}$ denotes the logarithmic average of the Eliashberg spectral function [Eq. (5)]. $\lambda$ is the EPC constant. Superconducting $T_c$ is calculated by using the McMillian-Allen-Dynes formula with $\mu^* =$  0.13. }
\begin{center}
\begin{tabular*}{8cm}{@{\extracolsep{\fill}} ccccccc}
\hline \hline
$\epsilon$ & $a$ & $c$ & $N(0)$ & $\omega_{log}$ & $\lambda$ & $T_c$ \\
(\%) &({\AA}) & ({\AA}) & (states/eV) & (cm$^{-1}$) &   & (K) \\
\hline
-4 & 4.958 & 5.282 & 0.96 & 161.1 & 0.590 & 3.51 \\
-2 & 5.061 & 5.222 & 1.02 & 158.0 & 0.614 & 4.05 \\
0 & 5.164 & 5.164 & 1.08 & 155.9 & 0.653 & 4.84 \\
2 & 5.267 & 5.108 & 1.12 & 147.8 & 0.727 & 6.29 \\
4 & 5.371 & 5.060 & 1.13 & 134.9 & 0.818 & 7.67 \\
\hline
\hline
\end{tabular*}
\end{center}
\end{table}

\begin{figure}[!t]
\includegraphics[angle=0,scale=0.52]{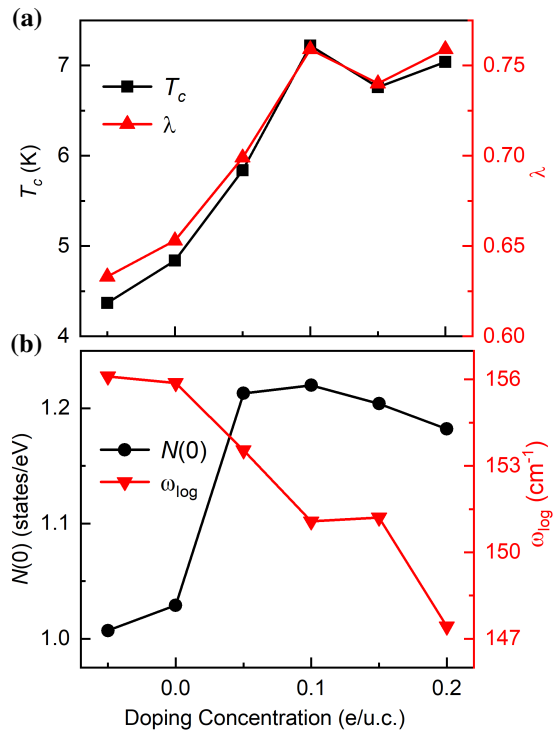}
\caption{Charge doping effect on the superconducting properties of bulk LaO. (a) Calculated superconducting $T_c$ (black squares) and EPC constant $\lambda$ (red triangles) as well as (b) calculated electronic density of states at the Fermi level $N(0)$ (black dots) and logarithmically averaged phonon frequency $\omega_{log}$ (red triangles), respectively, as functions of doping concentration.}
\label{fig4}
\end{figure}

\begin{figure*}[!t]
\includegraphics[angle=0,scale=0.5]{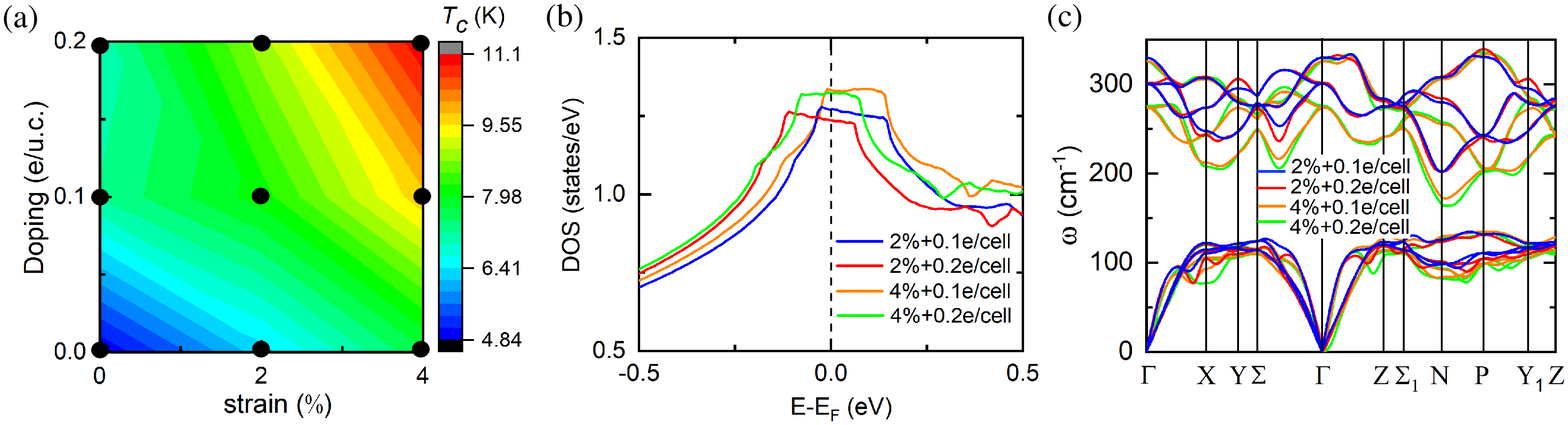}
\caption{ (a) Superconducting $T_c$ of LaO under the synergistic effect of tensile strain and electron doping. The nine black dots are the specific calculation results.  (b) Total electronic DOS and (c) phonon spectra of bulk LaO under various tensile strains and electronic doping concentrations.}
\label{fig5}
\end{figure*}

We further study the effects of lattice strain and charge doping on the superconducting properties of LaO. The in-plane biaxial strains from $-4\%$ (compressive strain) to $4\%$ (tensile strain) have been considered, which are accessible experimentally. From Table I, we can see that the compressive (tensile) in-plane strain causes the expansion (shrink) of the $c$-axis lattice, indicating that LaO undergoes a structural transition from the cubic phase to the tetragonal phase \cite{monoxide4}. These structural changes have significant impact on the electronic states of LaO. For example, the electronic DOS at the Fermi level decreases under the compressive strain but increases with the tensile strain (Table I), the latter of which would allow more electronic states to couple with the phonons. As the tensile strain increases, the phonon modes generally shift toward lower frequencies, especially where the EPC is strong (see Fig. \ref{fig7} in Appendix A); concomitantly, the total EPC strength $\lambda$ is lifted (Table I). The calculated superconducting $T_c$ reaches 7.67 K at $\epsilon=4\%$, which is enhanced by $58.5\%$ compared with the strain-free one (4.84 K). It is noteworthy that the simulated strain effect on superconductivity is in good agreement with the behavior of the measured $T_c$ of LaO thin films grown on different substrates (YAlO$_3$, LaAlO$_3$, LaSrAlO$_4$)\cite{monoxide4}.

The influence of charge doping on the superconductivity in LaO is also examined. Here, we simulate the charge doping by directly adding electrons into or removing electrons from the system, together with a compensating uniform charge background of opposite sign to maintain the charge neutrality. Considering that there is a hump in the DOS right above the Fermi level [see Fig. \ref{fig2}(b)], we infer that the electron doping is more conducive to enhancing EPC and $T_c$ than the hole doping. This is confirmed by the calculated values of $T_c$: 4.37 K for 0.05 holes/cell doping vs. 5.84 K for 0.05 electrons/cell doping [Fig. \ref{fig4}(a)]. We then focus on the case of electron doping in a range from 0.05 to 0.2 electrons/cell. As seen from Fig. \ref{fig4}(b), with the electron doping the electronic DOS at the Fermi level $N(0)$ first increases dramatically and then decreases gradually. Meanwhile, the calculated $T_c$ initially lifts up sharply with the electron doping, reaching a value of 7.22 K, and then varies modestly [see Fig. \ref{fig4}(a)].
The above results indicate that both the tensile lattice strain and the pure electron doping can enhance the superconducting $T_c$ of LaO.

We further investigate the synergistic effect of the tensile strain and the charge doping on the superconductivity in LaO. Figure \ref{fig5}(a) shows the distribution of $T_c$ under different lattice strains and doping concentrations, in which nine black dots mark the actual points we have calculated. Compared with the $T_c$ obtained under pure tensile strain or pure electron doping (Table I and Fig. \ref{fig4}), the $T_c$ increases significantly with their combined effects and reaches a value of 11.11 K at 0.2 electrons/cell doping and $4\%$ tensile strain, more than twice that of pure LaO. According to Fig. \ref{fig5}(b), the electronic DOS under $4\%$ tensile strain and 0.2 electrons/cell doping is enhanced, which is beneficial to the superconductivity. From the phonon spectra shown in Fig. \ref{fig5}(c), we can see that when the doping concentration increases from 0.1 to 0.2 electrons/cell, the phonon spectra appears obviously softening and would be further softened with the tensile strain increasing from $2\%$ to $4\%$, which would lead to the enhancements of EPC and $T_c$. In short, the significant boost in $T_c$ could be attributed to the softening of phonon spectrum and the increase in the electronic DOS at the Fermi level.


\begin{figure}[!b]
\includegraphics[angle=0,scale=0.52]{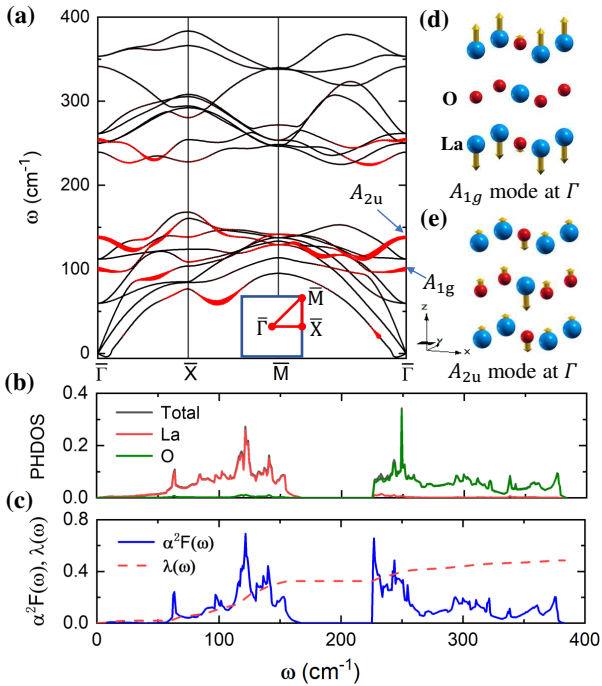}
\caption{(Color online) (a) Phonon spectrum of trilayer LaO. The size of red dot represents the EPC strength $\lambda_{{\bf q}\nu}$. (b) Total and projected phonon DOS. (c) Eliashberg spectral function $\alpha^2F(\omega)$ (blue line) and integrated $\lambda(\omega)$ (red dashed line). Vibrational patterns of (d) $A_{1g}$ and (e) $A_{2u}$ phonon modes at the $\overline{\Gamma}$ point with strong EPC. Yellow arrows and their lengths denote the direction and amplitude of atomic vibrations, respectively.}
\label{fig6}
\end{figure}

Besides the strain and charge doping effects, the dimensionality is another important factor that could influence the superconductivity of materials. We have examined whether the LaO thin film is still stable and superconducting down to the atomic-layer thickness. The calculated phonon spectra of monolayer and bilayer LaO thin films show that they both have large imaginary frequencies around the $\overline{X}$ point, indicating their dynamical instabilities. In contrast, there is no imaginary frequency in the whole BZ for the trilayer LaO film except a tiny one near the $\overline{\Gamma}$ point [Fig. \ref{fig6}(a)]. Instead of a sign of structure instability, it may result from the numerical difficulties in the accurate calculation of rapid decaying interatomic force\cite{artifact,high-order}, which was also found in the phonon spectra of borophene \cite{borophene}, germanene \cite{germanene}, and buckled arsenene \cite{arsenene1,arsenene2}.

We then focus on the stable trilayer LaO film. The orbital-resolved band structure shows that trilayer LaO is a metal with the states at the Fermi level mainly contributed by La 5$d$ orbitals (see Fig. \ref{fig8} in Appendix B). Similar to bulk LaO, the phonon spectrum of trilayer LaO can be divided into two regions [Figs. \ref{fig6}(a) and \ref{fig6}(b)]: a low-frequency branch (0 $\sim$ 170 cm$^{-1}$) with main composition of the La vibrations and a high-frequency branch (220 $\sim$ 400 cm$^{-1}$) mainly with the O vibrations. The mode- and momentum-resolved EPC strength [red dots in Fig. \ref{fig6}(a)] labels the phonon modes with large EPC. We select two typical phonon modes at the $\overline{\Gamma}$ point and plot their vibration patterns in Figs. \ref{fig6}(d) and \ref{fig6}(e), respectively. One is an $A_{1g}$ mode, where the outermost LaO layers show opposite vibrations along the $z$ direction and the middle layer remains static; meanwhile, each outermost LaO layer involves the in-phase vibrations of La and O atoms. The other is an $A_{2u}$ mode, which involves the out-of-phase vibrations of La and O atoms along the $z$ direction in each LaO layer. The calculated total EPC constant $\lambda$ of trilayer LaO film is 0.488 and the corresponding $T_c$ is 1.4 K. These results indicate that LaO is still stable and superconducting down to the trilayer thickness.

\section{CONCLUSION}

In summary, based on the first-principles electronic structure calculations, we show the conventional electron-phonon coupling mechanism for the superconductivity in rock-salt structural LaO. The calculated superconducting $T_c$ of 4.84 K is in good accordance with the previous measurements\cite{monoxide4}. According to our calculations, the EPC is mainly contributed by the low-frequency phonon modes of La atoms. Moreover, both the tensile lattice strain and the pure electron doping can boost the $T_c$, where the former is consistent with the previous observations on LaO films prepared on different substrates\cite{monoxide4} and the latter waits for future experimental verification. It is worth noting that when the doping concentration reaches 0.2 electrons/cell and the tensile strain reaches $4\%$, the $T_c$ is significantly enhanced to 11.11 K, indicating that the superconductivity of LaO can be significantly enhanced under the synergistic effect of electron doping and tensile strain. Interestingly, our calculations also suggest that the superconductivity ($T_c$ $\sim$ 1.4 K) still exists when the LaO film is reduced to trilayer thickness. Our works call for more effort to explore superconductivity in rare-earth compounds and their stacking structures \cite{laofese} as well as to study the related superconducting mechanism.

\begin{acknowledgments}

This work was supported by the National Key R\&D Program of China (Grant No. 2017YFA0302903 and No. 2019YFA0308603), the National Natural Science Foundation of China (Grant No. 11934020 and No. 11774424), the Beijing Natural Science Foundation (Grant No. Z200005), the CAS Interdisciplinary Innovation Team, the Fundamental Research Funds for the Central Universities, and the Research Funds of Renmin University of China (Grant No. 19XNLG13). Computational resources were provided by the Physical Laboratory of High Performance Computing at Renmin University of China.

\end{acknowledgments}

\begin{appendix}

\begin{figure}[]
\includegraphics[angle=0,scale=0.34]{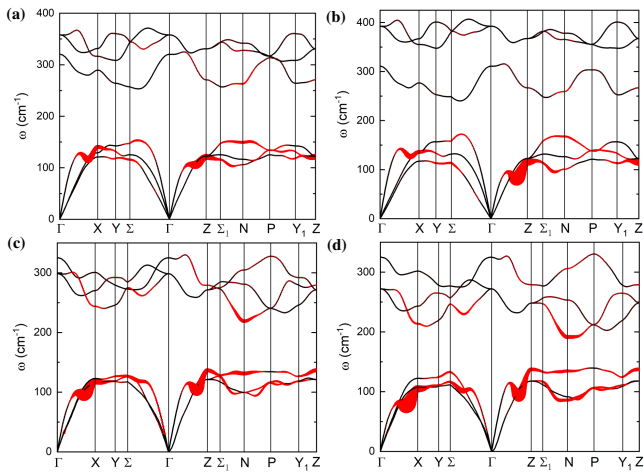}
\caption{Phonon spectrum of bulk LaO under different biaxial strains. The size of red dots on black lines represent the EPC strength $\lambda_{{\bf q}\nu}$. (a) $-2\%$ compressive strain; (b) $-4\%$ compressive strain; (c) $2\%$ tensile strain; (d) $4\%$ tensile strain.}
\label{fig7}
\end{figure}

\section{Phonon spectra and EPC strength of bulk LaO under different biaxial strains}

\begin{figure}[]
\includegraphics[angle=0,scale=0.43]{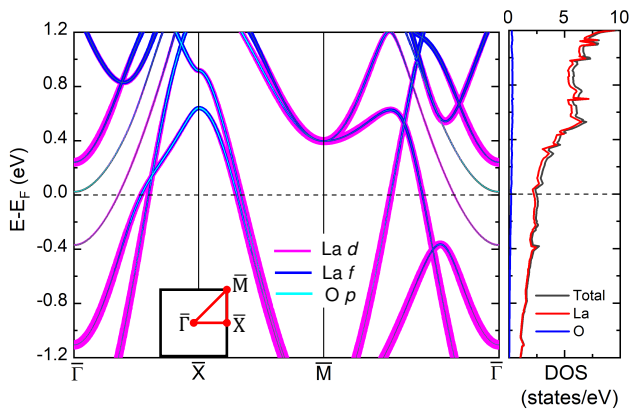}
\caption{Electronic structure of trilayer LaO. The left panel is the orbital-resolved band structure along the high-symmetry paths of the BZ. The bands contributed to by La 5$d$, 4$f$, and O 2$p$ orbitals are displayed with pink, blue, and cyan lines, respectively. The widths of lines are proportional to the corresponding orbital weights. The right panel shows the total and projected density of states for trilayer LaO. The Fermi level is set to zero.}
\label{fig8}
\end{figure}

To explore the effect of lattice stain on the superconductivity in bulk LaO, we calculated the phonon spectrum and the EPC strength under different strains ranging from $-4\%$ to $4\%$. As shown in Fig. \ref{fig7}, the phonon modes overall shift toward higher (lower) frequencies with the compressive (tensile) strain.

\section{Electronic band structure, total and projected DOS of trilayer LaO thin film}

Figure \ref{fig8} shows the electronic structures of trilayer LaO. There are three bands crossing the Fermi level $E_F$, indicating its metallic nature. According to the orbital-resolved band structure, the bands around the Fermi energy level are mainly composed of La 5$d$ orbitals while the weights of La 4$f$ orbitals and O 2$p$ orbitals are relatively lower.

\end{appendix}


\begin{thebibliography}{}

\bibitem{tc1969}R. W. Green, E. O. Thorland, J. Croat, and S. Legvold, J. Appl. Phys. {\bf 40}, 3161 (1969).
\bibitem{exp2014}V. Babizhetskyy, O. Jepsen, R. K. Kremer, A. Simon, B. Ouladdiaf, and A. Stolovits, J. Phys.: Condens. Matter {\bf 26}, 025701 (2014).
\bibitem{the2015}H. M. T\"{u}t\"{u}nc\"{u} and G. P. Srivastava, J. Appl.  Phys. {\bf 117}, 153902 (2015).
\bibitem{11tc1}A. L. Giorgi, E. G. Szklarz, M. C. Krupka, and N. H. Krikorian, J. Less-Commom Met. {\bf 17}, 121 (1969).
\bibitem{11tc2}A. L. Giorgi, E. G. Szklarz, N. H. Krikorian, and M. C. Krupka, J. Less-Commom Met. {\bf 22}, 131 (1970).
\bibitem{11tc3}T. L. Francavilla and F. L. Carter, Phys. Rev. {\bf 14}, 128 (1976).
\bibitem{13tc1}A. Simon and Th. Gulden, Z. Anorg. Allg. Chem. {\bf 630}, 2191 (2004).
\bibitem{13tc2}J. S. Kim, R. K. Kremer, O. Jepsen, and A. Simon, Curr. Appl. Phys. {\bf 6}, 897 (2006).
\bibitem{13tc3}J. S. Kim, W. Xie, R. K. Kremer, V. Babizhetskyy, O. Jepsen, A. Simon, K. S. Ahn, B. Raquet, H. Rakoto, J.-M. Broto, and B. Ouladdiaf, Phys. Rev. B {\bf 76}, 014516 (2007).
\bibitem{multip1}K. Sugawara, T. Sato, S. Souma, T. Takahashi, and A. Ochia, Phys. Rev. B {\bf 76}, 132512 (2007).
\bibitem{multip2}S. Kuroiwa, Y. Saura, J. Akimitsu, M. Hiraishi, M. Miyazaki, K. H. Satoh, S. Takeshita, and R. Kadono, Phys. Rev. Lett. {\bf 100}, 097002 (2008).
\bibitem{multip3}A. Poto\v{c}nik, P. Jegli\v{c}, K. Kobayashi, K. Kawashima, S. Kuchida, J. Akimitsu, and D. Ar\v{c}on, Phys. Rev. B {\bf 90}, 104507 (2014).
\bibitem{review2017}M. Smidman, M. B. Salamon, H. Q. Yuan, and D. F. Agterberg, Rep. Prog. Phys. {\bf 80}, 036501 (2017).
\bibitem{XMR1}F. F. Tafti, Q. D. Gibson, S. K. Kushwaha, N. Haldolaarachchige, and R. J. Cava, Nat. Phys. {\bf 12}, 272 (2016).
\bibitem{XMR2}L.-K. Zeng, R. Lou, D.-S. Wu, Q.N. Xu, P.-J. Guo, L.-Y. Kong, Y.-G. Zhong, J.-Z. Ma, B.-B. Fu, P. Richard, P. Wang, G. T. Liu, L. Lu, Y.-B. Huang, C. Fang, S.-S. Sun, Q. Wang, L. Wang, Y.-G. Shi, H. M. Weng \textit{et al}., Phys. Rev. Lett. {\bf 117}, 127204 (2016).
\bibitem{XMR3}F. Fallah Tafti, Q. Gibson, S. Kushwaha, J. W. Krizan, N. Haldolaarachchige, and R. J. Cava, Proc. Nat. Acad. Sci. U.S.A. {\bf 113}, E3475 (2016).
\bibitem{XMR4}S.-S. Sun, Q. Wang, P.-J. Guo, K. Liu, and H.-C. Lei, New J. Phys. {\bf 18}, 082002 (2016).
\bibitem{LaSbTc}M. Zhang, X. Wang, A. Rahman, R. Dai, Z. Wang, and Z. Zhang, Phys. Rev. B {\bf 101}, 064106 (2020).
\bibitem{LaBiTc}F. F. Tafti, M. S. Torikachvili, R. L. Stillwell, B. Baer, E. Stavrou, S. T. Weir, Y. K. Vohra, H.-Y. Yang, E. F. McDonnell, S. K. Kushwaha, Q. D. Gibson, R. J. Cava, and J. R. Jeffries, Phys. Rev. B {\bf 95}, 014507 (2017).
\bibitem{jfzhang2020}J.-F. Zhang, P.-J. Guo, M. Gao, K. Liu, and Z.-Y. Lu, Phys. Rev. B {\bf 101}, 155139 (2020).
\bibitem{LaH10exp1}M. Somayazulu, M. Ahart, A. K. Mishra, Z. M. Geballe, M. Baldini, Y. Meng, V. V. Struzhkin, and R. J. Hemley, Phys. Rev. Lett. {\bf 122}, 027001 (2019).
\bibitem{LaH10exp2}A. P. Drozdov, P. P. Kong, V. S. Minkov, S. P. Besedin, M. A. Kuzovnikov, S. Mozaffari, L. Balicas, F. F. Balakirev, D. E. Graf, V. B. Prakapenka, E. Greenberg, D. A. Knyazev, M. Tkacz, and M. I. Eremets, Nature (London) {\bf 569}, 528 (2019).
\bibitem{LaH10the1}F. Peng, Y. Sun, C. J. Pickard, R. J. Needs, Q. Wu, and Y. Ma, Phys. Rev. Lett. {\bf 119}, 107001 (2017).
\bibitem{LaH10the2}H. Liu, I. I. Naumov, R. Hoffmann, N. W. Ashcroft, and R. J. Hemley, Proc. Natl. Acad. Sci. USA {\bf 114}, 6990 (2017).
\bibitem{LaBaCuO}J. G. Bednorz and K.A. M\"{u}ller, Z. Phys. B: Condens. Matter. {\bf 64}, 189 (1986).
\bibitem{CuO21}T. Jarlborg and A. Bianconi, Phys. Rev. B {\bf 87}, 054514 (2013).
\bibitem{CuO22}W. M. Li, J. F. Zhao, L. P. Cao, Z. Hu, Q. Z. Huang, X. C. Wang, Y. Liu, G. Q. Zhao, J. Zhang, Q. Q. Liu, R. Z. Yu, Y. W. Long, H. Wu, H. J. Lin, C. T. Chen, Z. Li, Z. Z. Gong, Z. Guguchia, J. S. Kim, G. R. Stewart \textit{et al}., Proc. Natl. Acad. Sci. U.S.A. {\bf 116}, 12156 (2019).
\bibitem{LaOFeAs}Y. Kamihara, T. Watanabe, M. Hirano, and H. Hosono, J. Am. Chem. Soc. {\bf 130}, 3296 (2008).
\bibitem{FeAs1}R. Caivano, M. Fratini, N. Poccia, A. Ricci, A. Puri, Z.-A. Ren, X.-L. Dong, J. Yang, W. Lu, Z.-X. Zhao, L. Barba, and A. Bianconi, Super. Sci. and Tech. {\bf 22}, 014004 (2009).
\bibitem{FeAs2}D. J. Singh and M. H. Du, Phys. Rev. Lett. {\bf 100}, 237003 (2008).
\bibitem{FeAs3}I. I. Mazin, D. J. Singh, M. D. Johannes, and M. H. Du, Phys. Rev. Lett. {\bf 101}, 057003 (2008).
\bibitem{instablity}S. Zhang and R. Xiao, J. Appl. Phys. {\bf 83}, 3842 (1998).
\bibitem{monoxide1}K. Kaminaga, R. Sei, K. Hayashi, N. Happo, H. Tajiri, D. Oka, T. Fukumura, and T. Hasegawa, Appl. Phys. Lett. {\bf 108}, 122102 (2016).
\bibitem{monoxide2}Y. Uchida, K. Kaminaga, T. Fukumura, and T. Hasegawa, Phys. Rev. B {\bf 95}, 125111 (2017).
\bibitem{monoxide3}K. Kaminaga, D. Oka, T. Hasegawa, and T. Fukumura, ACS Omega {\bf 3}, 12501 (2018).
\bibitem{monoxide4}K. Kaminaga, D. Oka, T. Hasegawa, and T. Fukumura, J. Am. Chem. Soc. {\bf 140}, 6754 (2018).
\bibitem{monoxide5}T. Yamamoto, K. Kaminaga, D. Saito, D. Oka, and T. Fukumura, Appl. Phys. Lett. {\bf 114}, 162104 (2019).
\bibitem{dft1}P. Hohenberg and W. Kohn, Phys. Rev. {\bf 136}, B864 (1964).
\bibitem{dft2}W. Kohn and L. J. Sham, Phys. Rev. {\bf 140}, A1133 (1965).
\bibitem{dfpt1}S. Baroni, S. de Gironcoli, A. Dal Corso, and P. Giannozzi, Rev. Mod. Phys. {\bf 73}, 515 (2001).
\bibitem{dfpt2}F. Giustino, Rev. Mod. Phys. {\bf 89}, 015003 (2017).
\bibitem{QE1}P. Giannozzi, S. Baroni, N. Bonini, M. Calandra, R. Car, C.
Cavazzoni, D. Ceresoli, G. L. Chiarotti, M. Cococcioni, I. Dabo \textit{et al}., J. Phys.: Condens. Matter {\bf 21}, 395502 (2009).
\bibitem{uspp}A. M. Rappe, K. M. Rabe, E. Kaxiras, and J. D. Joannopoulos, Phys. Rev. B {\bf 41}, 1227 (1990).
\bibitem{pslibrary1}A. Dal Corso, Comput. Mater. Sci. {\bf95}, 337 (2014).
\bibitem{pslibrary2}https://www.quantum-espresso.org/pseudopotentials.
\bibitem{PBE}J. P. Perdew, K. Burke, and M. Ernzerhof, Phys. Rev. Lett. {\bf 77}, 3865 (1996).
\bibitem{BFGS}S. R. Billeter, A. Curioni, and W. Andreoni, Comput. Mater. Sci. {\bf 27}, 437 (2003).
\bibitem{Eliashberg}G. M. Eliashberg, Zh. Eksp. Teor. Fiz. {\bf 38}, 966 (1960) [Sov. Phys. JETP {\bf 11}, 696 (1960)].
\bibitem{McMillan1}P. B. Allen, Phys. Rev. B {\bf 6}, 2577 (1972).
\bibitem{miu1}K.-H. Lee, K. J. Chang, and M. L. Cohen, Phys. Rev. B {\bf 52}, 1425 (1995).
\bibitem{miu2}C. F. Richardson and N. W. Ashcroft, Phys. Rev. Lett. {\bf 78}, 118 (1997).
\bibitem{LaS}S. Sankaralingam, S. Mathijaya, G. Pari, and R. Asokamani, Phys. Stat. Sol. (b) {\bf 174}, 435 (1992).
\bibitem{beforexp}J. M. Leger, N. Yacoubi, and J. Loriers, J. Solid State Chem. {\bf 36}, 261 (1981).
\bibitem{artifact}H. \c{S}ahin, S. Cahangirov, M. Topsakal, E. Bekaroglu, E. Akturk, R. T. Senger, and S. Ciraci, Phys. Rev. B {\bf 80}, 155453 (2009).
\bibitem{high-order}The calculation including high-order interatomic force constants contains the nonlinear effect, which can be helpful to obtain more accurate results but is beyond the scope of the present paper.
\bibitem{borophene}M. Gao, Q.-Z. Li, X.-W. Yan, and J. Wang, Phys. Rev. B {\bf95}, 024505 (2017).
\bibitem{germanene}S. Cahangirov, M. Topsakal, E. Akt\"{u}rk, H. \c{S}ahin, and S. Ciraci, Phys. Rev. Lett. {\bf 102}, 236804 (2009).
\bibitem{arsenene1}C. Kamal and M. Ezawa, Phys. Rev. B {\bf 91}, 085423 (2015).
\bibitem{arsenene2}X. Kong, M. Gao, X.-W. Yan, Z.-Y. Lu, and T. Xiang, Chin. Phys. B {\bf 27}, 046301 (2018).
\bibitem{laofese}X.-L. Qiu, B.-C. Gong, H.-C. Yang, Z.-Y. Lu, and K. Liu, Phys. Rev. B {\bf 103}, 035143 (2021).


\end{thebibliography}
\end{document}